\documentclass[aps,pre,amsmath,amssymb,twocolumn,nopacs,floatfix,superscriptaddress]{revtex4-2}

\usepackage{mathtools}
\usepackage{amsmath}
\usepackage{amssymb}
\usepackage{bm}
\usepackage{graphicx}
\usepackage{xcolor}
\usepackage{epstopdf}
\usepackage{subfig}
\usepackage{subfloat}
\usepackage{dcolumn}
\usepackage{bm}

\usepackage{amsfonts,amssymb,amsmath,array}
\usepackage{graphicx}
\usepackage{amsbsy}
\usepackage{color}
\usepackage{hyperref} 
\usepackage{enumerate}
\usepackage{mathtools}
\usepackage{lipsum}

\begin{document}

\title{Thermodynamic precision of a chain of motors: \\the difference between phase and noise correlation}

\author{G. Costantini}
\affiliation{Department of Physics, University of Rome Sapienza, P.le Aldo Moro 2, 00185, Rome, Italy}\affiliation{Institute for Complex Systems - CNR, P.le Aldo Moro 2, 00185, Rome, Italy}

\author{A. Puglisi}
\affiliation{Department of Physics, University of Rome Sapienza, P.le Aldo Moro 2, 00185, Rome, Italy}\affiliation{Institute for Complex Systems - CNR, P.le Aldo Moro 2, 00185, Rome, Italy}
\affiliation{INFN, University of Rome Tor Vergata, Via della Ricerca Scientifica 1, 00133, Rome, Italy}

%\date{\today}
\begin{abstract}
Inspired by recent experiments on fluctuations of the flagellar beating in sperms and C. reinhardtii, we investigate the precision of  phase fluctuations in a system of nearest-neighbour-coupled molecular motors. We model the system as a Kuramoto chain of oscillators with coupling constant $k$ and noisy driving. The precision $p$ is a Fano-factor-like observable which obeys the Thermodynamic Uncertainty Relation (TUR), that is an upper bound related to dissipation. We first consider independent motor noises with diffusivity $D$: in this case the precision goes as $k/D$, coherently with the behavior of spatial order. The minimum observed precision is that of the uncoupled oscillator $p_{unc}$, the maximum observed one is $Np_{unc}$, saturating the TUR bound. Then we consider driving noises which are spatially correlated, as it may happen in the presence of some direct coupling between adjacent motors. Such a  spatial correlation in the noise does not reduce evidently the degree of spatial correlation in the chain, but sensibly reduces the maximum attainable precision $p$, coherently with experimental observations. The limiting behaviors of the precision, in the two opposite cases of negligible interaction and strong interaction, are well reproduced by the precision of the single chain site $p_{unc}$ and the precision of the center of mass of the chain $N_{eff} p_{unc}$ with $N_{eff}<N$: both do {\em not} depend on the degree of interaction in the chain, but $N_{eff}$ decreases with the correlation length of the motor noises.
\end{abstract}

\maketitle

\section{Introduction}

Biological processes continuously involve energy dissipation and therefore are inherently out of thermodynamic equilibrium~\cite{seifertrev}. A paramount case is that of molecular motors that typically convert chemical energy stored in ATP into mechanical energy to move a cargo or actuate the deformation of parts of a cell~\cite{peliti2021stochastic}. The cooperation of several molecular motors, for instance during the excitation of travelling waves in flagella or cilia,  is still an open problem, that can get new insight from the application of theoretical results in non-equilibrium statistical physics~\cite{julicher1997modeling}: in particular, mesoscopic fluctuations - under the lens of stochastic thermodynamics - can provide access to the underlying microscopic mechanisms producing such fluctuations~\cite{battle2016broken,ferretta2023thermal}. This has been recently shown in an experiment with mammalian sperms, where the fluctuations of the tail beating wave have been analysed and found to be more irregular than expected~\cite{maggi2023thermodynamic}. The amount of noise in a driven system, such as the beating of a sperm tail, is bounded from below by the amount of dissipation, through the celebrated Thermodynamic Uncertainty Relations (TURs), which we resume here~\cite{barato2015thermodynamic,gingrich2016dissipation,horowitz2020thermodynamic}. 

The simplest TUR mainly concerns systems with a time-integrated current $\theta(t)$ that steadily grows and fluctuates, for instance - at not too small times - its average grows as $\langle \theta(t) \rangle=Jt$ and its variance grows as $\textrm{Var}(\theta)=2 \sigma t$
%\begin{equation} \label{basic}
%\theta(t) \approx Jt + \sqrt{2D t }\mathcal{N},
%\end{equation}
%with $\mathcal{N}$ a random variable with zero average  and finite variance, 
where $J$ and $\sigma$ are the constant average current rate and diffusivity, respectively. This minimal picture is qualitatively meaningful for 1) the position of Brownian motion  with an external constant force; 2) a current in a Markov process with broken detailed balance, such as in several models for molecular motors. Of course it encompasses all stochastic variables which admit a large deviation principle. In those systems it is possible to define a precision
\begin{equation}
p=\frac{J^2}{\sigma}
\end{equation}
which 1) in a steady state is a constant, 2) has the dimension of an inverse time, and 3) its inverse $t^*=1/p$ represents the time that separates the small time regime dominated by fluctuations and the large time regime dominated by the average growth. A very succinct summary of the so-called thermodynamic uncertainty relation (TUR) states that
\begin{equation} \label{eq:basictur}
p \le  \frac{\dot S_p}{k_B }= \frac{\dot  W}{k_B T}
\end{equation}
where $\dot S_p$ is the entropy production rate in the system and the last equality holds for systems driven out-of-equilibrium by forces that inject power $\dot W$ and in contact with a thermal bath at temperature $T$~\cite{barato2015thermodynamic,gingrich2016dissipation,horowitz2020thermodynamic}. It has also been demonstrated that there is a diffusion process with drift, whose large deviation rate constrains the large deviation rate of the observed current, with a diffusion coefficient that saturates the TUR~\cite{gingrich2017fundamental,nardini2018process}.

In a recent experiment the maximum “thermodynamic precision” of the sperm’s tail has been measured to be $p \sim \frac{1}{N}\frac{\dot  W}{k_B T}$ where $N \sim 10^5$ is a good approximation of the number of molecular motors (dynein) present along the axoneme that actuates the tail~\cite{maggi2023thermodynamic}. Additionally, it has also been seen that reducing $\dot W$ (by slowly depleting the sperm's nutrients) leads to a proportional reduction of $p$ leaving the $\frac{1}{N}$ factor substantially unaltered. This suggests that the sperm tail is behaving closely to a single dynein motor which, in fact, is semi-optimised ~\cite{hwang2018energetic} having a precision similar to that measured in the tail but an energy consumption $N$ times smaller.  So it seems that the current measured to define the sperm’s tail precision is proportional to the current that  can be measured to define the dynein’s precision: these two “Brownian clocks” are proportional including their fluctuations. A comparison with other~\cite{polin2009chlamydomonas,goldstein2009noise,goldstein2011emergence,wan2014rhythmicity,ma2014active,quaranta2015hydrodynamics} - indirect - measurements of $p$ with sperms in different setups and with C. reinhardtii (which swims using two flagella made of the same axonemal structure but with a smaller length) is coherent with the previous picture. All these observations could be explained by conjecturing a strong coupling of {\em fluctuations} in the dynamics of nearby molecular motors, which is reasonable by considering the high dense packing of motors in the axoneme, and by some direct evidence through electron microscopy~\cite{brokaw2009thinking,burgess1995rigor,goodenough1982substructure}. 

\begin{figure}
    \includegraphics[width=\columnwidth]{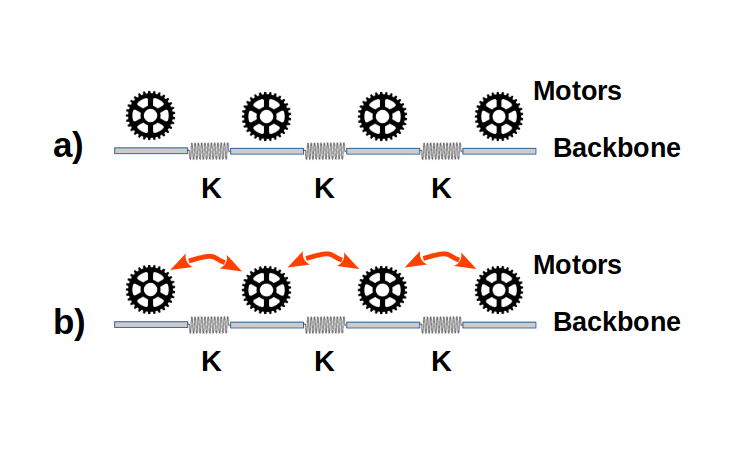}
    \caption{ \label{fig1} A scheme of the interpretation of the two models considered here. In both cases the model is described by the  configuration of the $N$ parts of the chain, here denoted as "backbone" (representing e.g. a biological flagellum) $\theta_i(t)$ and the configuration of the $N$ motors that actuate the chain. Top: only the sub-parts of the backbone are coupled through a stiffness $k$, while the motors are completely uncoupled. Bottom: an additional coupling is introduced which correlates the states of adjacent motors. }
\end{figure}

Here we address this problem on a more abstract and, possibly, general ground: in a meso-scopic system driven by many microscopic motors, can the direct coupling (spatial correlation) among fluctuations of the microscopic motors be detrimental for the precision of the meso-scopic system? We stress that models of multi-motor systems~\cite{julicher1997spontaneous,guerin2011bidirectional,guerin2011dynamical} already take into account an {\em indirect} coupling among the motors, also named cooperativity, which comes through the fact that each  motor is coupled to the state of the axonemal backbone in the point where the motor is attached, so that adjacent motors feel a similar "external field" and are - therefore - indirectly coupled to each other. Here we neglect the feedback effect of the backbone onto the motors and therefore we ignore the  indirect coupling between motors that would be mediated by the backbone. We only consider the direct coupling between adjacent parts of the backbone (through its elasticity $k$) and - in the second part of the paper - a direct coupling among the noises which represent a direct (e.g. mechanical) interaction among the underlying motors, see Fig.~\ref{fig1}.

We brievly riepilogate some papers which address questions similar but not identical to the present one. 
%Recent works on similar questions gave different (and inconsistent) answers~\cite{izumida2016energetics,lee2018thermodynamic,zhang2020energy,hong2020thermodynamic}.  These recent works were based upon mean field models of synchronization where all units interacted with all units. In these work not only the precision but also the energetic cost of systems made of multiple coupled motors is considered. In the present paper we impose the energetic cost to be independent from the other parameters. 
In~\cite{izumida2016energetics} the energetic cost of two oscillators coupled by a generic force is considered. It is shown that the potential part of the interaction always decrease dissipation, while the non-potential part may decrease or increase it. In the same spirit~\cite{zhang2020energy} considers coupled oscillators but with a more complicate coupling which happens to be purely non-potential, with the result of increasing the cost when interaction (and therefore order) increases. In ~\cite{lee2018thermodynamic} the effect of a conservative (Kuramoto) coupling on the TUR is considered, but the TUR is replaced by a "sub-system-TUR" which compares fluctuations of a sub-system with the energy dissipated by the same sub-system, a measure  which in many cases gives simply $1/N$ of the total dissipated energy. This amounts to say that the sub-system has a precision which is bounded by the same bound of the whole system. Finally~\cite{hong2020thermodynamic} considers the total thermodynamic cost in a mean-field Kuramoto model, somehow recalling and reproducing for a single model the general result of~\cite{izumida2016energetics}, where a potential coupling leads to reducing dissipation (when interaction and correlation increases), while no discussion of precision and TUR are given.

A general formalization of this problem is the following, see Fig.~\ref{fig1}. In all these models one has a set of phases $\theta_i(t)$ that represent a discretization (in space) of the backbone of the flagellum, i.e. each $i$-th phase represents some local observable (e.g. the local angle with respect to a fixed axis or the local curvature, etc.). The local ($i$-th) phase evolves under the effect of interaction and motors,  $\dot \theta_i(t)=F_i(t)$ with $F_i(t)=F_{i,int}(t)+F_{i,mot}(t)$: $F_{i,int}$ represents some non-linear couplings with adjacent parts of the backbone, e.g. $F_{i,int}(\theta_i,\theta_{i+1},\theta_{i-1})$ (the coupling is typically non-confining, so that phase slips are allowed), while $F_{i,mot}=\omega_i+\eta_i(t)$ represents the  coupling with the $i$-th molecular motor which has an average driving force $\omega_i$ and associated fluctuations $\eta_i(t)$. In this paper we consider $F_{i,int}$ to be conservative, i.e. derived from a potential, and of the Kuramoto type.

A typical precision measure in these models is related to a single current, the simplest case given by $\theta_i$ with a given $i$, i.e. in the steady state $p= \lim_{t \to\infty}\frac{2}{t}\langle \theta_i \rangle^2/\langle (\delta \theta_i)^2 \rangle$ with $\delta \theta_i=\theta(t)-\langle \theta \rangle$. The comparison of such a single current, in the spirit of the TUR, is with some measure of energy cost or dissipation: the total dissipation rate in the system can be decomposed as $\dot W=\sum_i \dot W_i $ where $\dot W_i=\langle \dot \theta_i \circ F_i \rangle$ is the energy dissipation rate of the single variable $\theta_i$, which is - in principle - influenced by couplings. The effect of conservative couplings (which is the case of Kuramoto models) is always non-increasing the total dissipation rate, and usually leaves it unchanged. Non-potential couplings can both increase or decrease the dissipation. We have verified that the total dissipation rate in our case is substantially unaltered by the value of $k$.

In Section~\ref{sec:model} we present the Kuramoto model with the two kinds of noise which we considered in our study. In Section~\ref{sec:uncor} the results with independent noises are discussed. In Section~\ref{sec:cor} we report the results with spatial correlations in the noise. Conclusions and perspective are drawn in Sec.~\ref{sec:concl} 

%The purpose of this new work is to understand if there is a reasonable range of validity of our conjecture (no big change in precision and energy cost) for a chain-like topology of interaction and (later) for other topologies which are more realistic for a flagellum.

\section{Model}
\label{sec:model}

We consider a chain of $N$ coupled oscillators:
\begin{equation}
\dot\theta_i = \omega_i -\frac{k}{2}[\sin(\theta_i-\theta_{i-1})+\sin(\theta_i-\theta_{i+1})]+\sqrt{2D}\eta_i(t)
\end{equation}
with $i=1..N$, $k \ge 0$ is the coupling constant, $D$ is the bare diffusivity parameter, $\eta_i$ a white noises with $\langle \eta_i \rangle = 0$ and two possible recipes for space correlations, the first is the uncorrelated case, discussed in Sec.~\ref{sec:uncor} the second is the correlated one, discussed in Sec.~\ref{sec:cor}:
\begin{subequations}
    \begin{align}
        &\langle \eta_i(t)\eta_j(t')\rangle=\delta_{ij}\delta(t-t')\\
               &\langle \eta_i(t)\eta_j(t')\rangle= \sum_{n=1}^{N} e^{-|i-n|/\Delta} e^{-|j-n|/\Delta}\delta(t-t') = \label{eq:cor}  \\
&\frac{1}{1-e^{-2/\Delta}}\Big\{e^{-|d|/\Delta}\Big[1+|d|+(1-|d|)e^{-2/\Delta}\Big]- \nonumber \\
&e^{-s/\Delta}-e^{-(2N-s+2)/\Delta}\Big\} \nonumber
    \end{align}
\end{subequations}
where $d\equiv j-i$ and $s\equiv j+i$. Formula~\eqref{eq:cor} is chosen for its easiest implementation, see Sec.~\ref{sec:cor}. When $N$ is large (already with $N=10^2$ this is true) and $s$ is of order $N$ (e.g. if $i$ is in the bulk of the chain) and $|d| \gg 1$, Eq.~\eqref{eq:cor} reduces to $\langle \eta_i(t)\eta_j(t')\rangle \approx |i-j|e^{-|i-j|/\Delta}$. In the present work we mainly consider uniform driving, $\omega_i=\omega$, with a few exceptions where the effect of a narrow distribution of $\omega_i$ is discussed.

 When $k=0$ we have exactly an overdamped Langevin equation  
 %are exactly in the situation of Eq.~\eqref{basic} 
 for each variable $\theta_i$ with $J=\omega_i$, then the uncoupled value of the local precision is $p_{unc,i}=\omega_i^2/D$ and in the uniform case we can define $p_{unc}=\omega^2/D$. It is immediately evident that in the uncoupled case the TUR, Eq.~\ref{eq:basictur} is {\em not} saturated, since $\dot W=N\omega^2$ and $k_B T=D$.

We recall that the Kuramoto chain, differently from the original mean field Kuramoto model, does not have a synchronization transition and a long-range order~\cite{strogatz1988phase,daido1988lower}, but short-range order can be revealed by suitable order parameter, as discussed later~\cite{giver2011phase,gutierrez2023nonequilibrium}.
%%%%%%%%%%%%%%%%%%%%%%%%%%%%%%

%%%%%%%%%%%%%%%%%%%%%%%%%%%%%%
\section{Results with independent noise}
\label{sec:uncor}

\begin{figure}
    \includegraphics[width=\columnwidth]{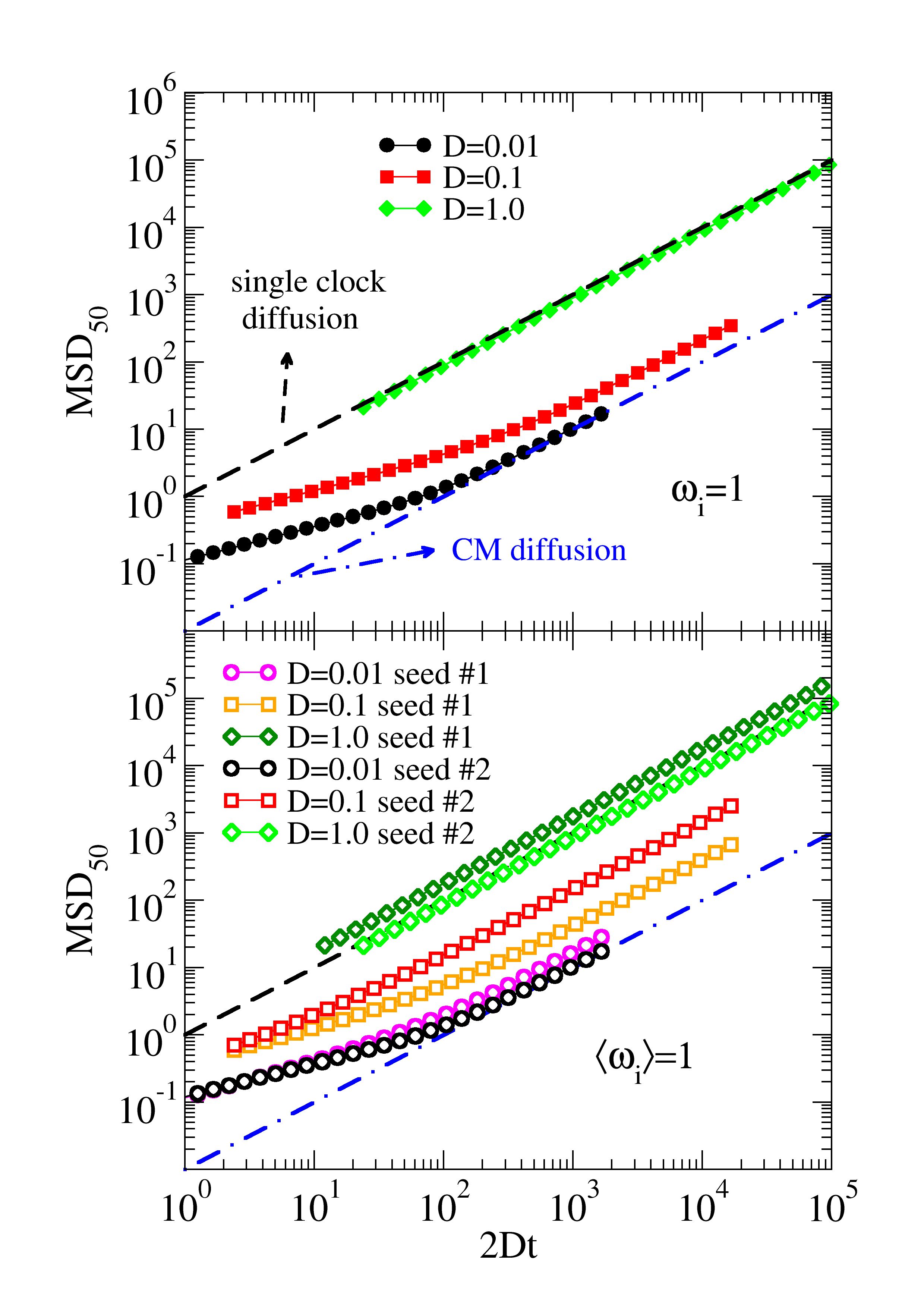}
    \caption{\label{fig:msd} Mean squared displacement as a function of rescaled time $2Dt$ for several choices of the parameters. The top graph shows cases with uniform driving $\omega_i=\omega$, the lower graph shows cases with a Gaussian distribution of $\omega_i$ with average $\omega$ and standard deviation  equal to the $5\%$ of $\omega$. The black dashed line represents the bare diffusion $\text{MSD} \sim 2D t$, the blue dot-dashed line represents the center-of-mass diffusion $\text{MSD} \sim 2(D/N) t$. Here $N=100$ as in the rest of the paper. }
\end{figure}

The aim of this section is to show the effect of chain short-range order upon the precision of phase fluctuations. 
We have simulated the Kuramoto chain for several choices of the parameters $D$, $k$, $\{\omega\}$ for long trajectories. From each long trajectory we focus on the central oscillator ($i=50$ in the middle of the chain of length $N=100$) in order to reduce the effects of the boundaries. We have verified that - apart from the oscillators very close to the boundaries, the results do not strongly depend upon $i$. We have computed time averages to get values for the average phase drift velocity $\langle \dot \theta_i \rangle$ and the mean squared displacement (MSD), which is shown for the site $i=50$ in Fig.~\ref{fig:msd}. The MSD is seen to reach the asymptotic normal regime $\text{MSD}_i \sim 2 D_i t$ after more or less long transients, depending on the parameters. In the case of uniform driving the asymptotic MSD is bounded from above by the raw diffusion coefficient $\sim 2 D t$  - reached for strong $D/k$ and from below by the center of mass diffusion $\sim 2(D/N) t$, reached for low $D/k$, as explained later. When the driving is not uniform the situation is similar but the bounds are not strict and one can observe strong differences between different choices of the values of $\omega_i$ within the same distribution.

\begin{figure} 
    \includegraphics[width=\columnwidth]{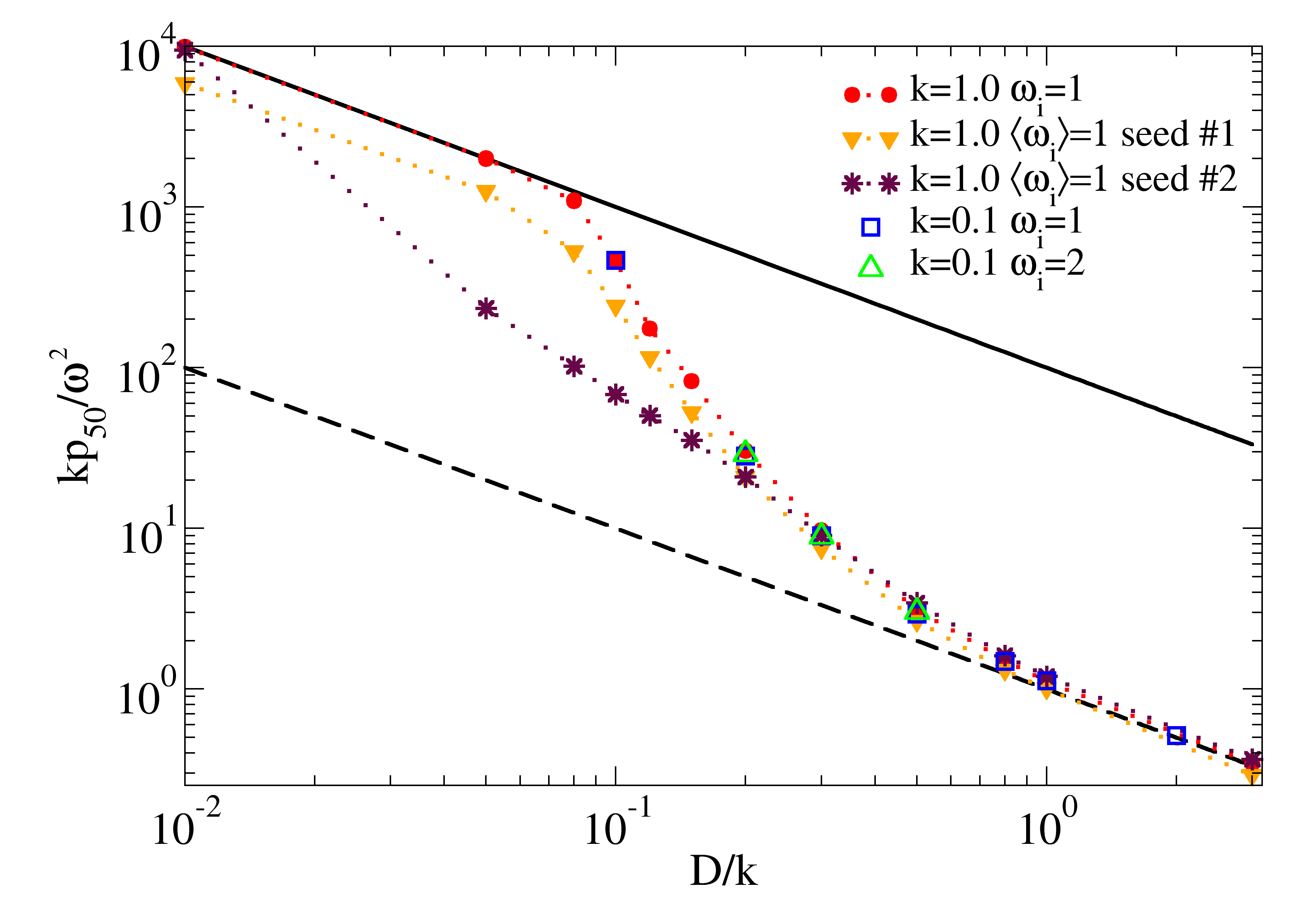}
    \caption{Precision, rescaled by $\omega^2/k$, of the site $i=50$ as a function of $D/k$, for several choices of the parameters, with uniform or non-uniform driving. The dashed line marks the inferior bound $1/D$ while the solid line marks the upper bound $N/D$. \label{fig:indnoise} }
\end{figure}

In all cases, however, it is possible to measure a diffusivity $D_i$ and compute the precision
\begin{equation}
p_i=\frac{\langle \dot \theta_i \rangle^2}{D_i}
\end{equation}
which is shown in Fig.~\ref{fig:indnoise}. For uniform $\omega_i=\omega$ we get a neat collapse by plotting $k p_i/\omega^2$ vs. $D/k$. The master curve shows two clear limiting behaviors, already anticipated by the two limiting behaviors of $D_i$
\begin{subequations}
    \begin{align}
        p_i \approx 1/D \;\;\;\; (D/k \gg 1) \\
        p_i \approx N/D \;\;\;\; (D/k \ll 1)
    \end{align}
\end{subequations}
The explanation for these two limiting behaviors is the following. When $D/k$ is large, the level of order is low (see later, where we characterize the degree of order in the chain), and therefore each oscillator is substantially independent from the others and follows a biased Brownian motion $\theta_i \sim \omega_i t+ W^D_t $ where $W^D_t$ is the Wiener process with diffusivity $D$.
On the contrary, when $D/k$ is small, there is a high level of order and all the oscillators fluctuate close to each other and, consequently, close to the average phase $\overline{\theta}=\frac{1}{N}\sum_{i=1}^N \theta_i$. We note that the average motion is not influenced by the internal interactions, i.e.
\begin{equation}
\overline{\dot\theta}=\overline{\omega}+ \frac{\sqrt{2D} }{N}\sum_{i=1}^N \eta_i =\overline{\omega}+ \sqrt{\frac{2D}{N}} \eta
\label{eq:cm}
\end{equation}
with $\eta$ a white noise with $1$ amplitude. The mean squared displacement of the single phase, asymptotically, cannot be slower or faster than the mean squared displacement of the average phase, i.e. $D_i \to D/N$ and this explains the behavior $p_i \approx N/D$ at small $D/k$.
% \dot\overline\theta=\overline\omega + \sqrt{2D} \frac{1}{N}\sum_{i=1}^N \eta_i   

In the case of uniform phase velocity $\omega_i=\omega=\overline\omega$, the rate of dissipated energy is trivially scaling with $N$, i.e. $w=\sum_i \langle \omega_i \dot \theta_i \rangle = N \omega^2$.  Therefore the maximum precision imposed by the TUR is $p_{max}=w/D=N\omega^2/D$ and this maximum is exactly saturated in the strong coupling limit, $D/k \ll 1$.

\begin{figure} 
    \includegraphics[width=\columnwidth]{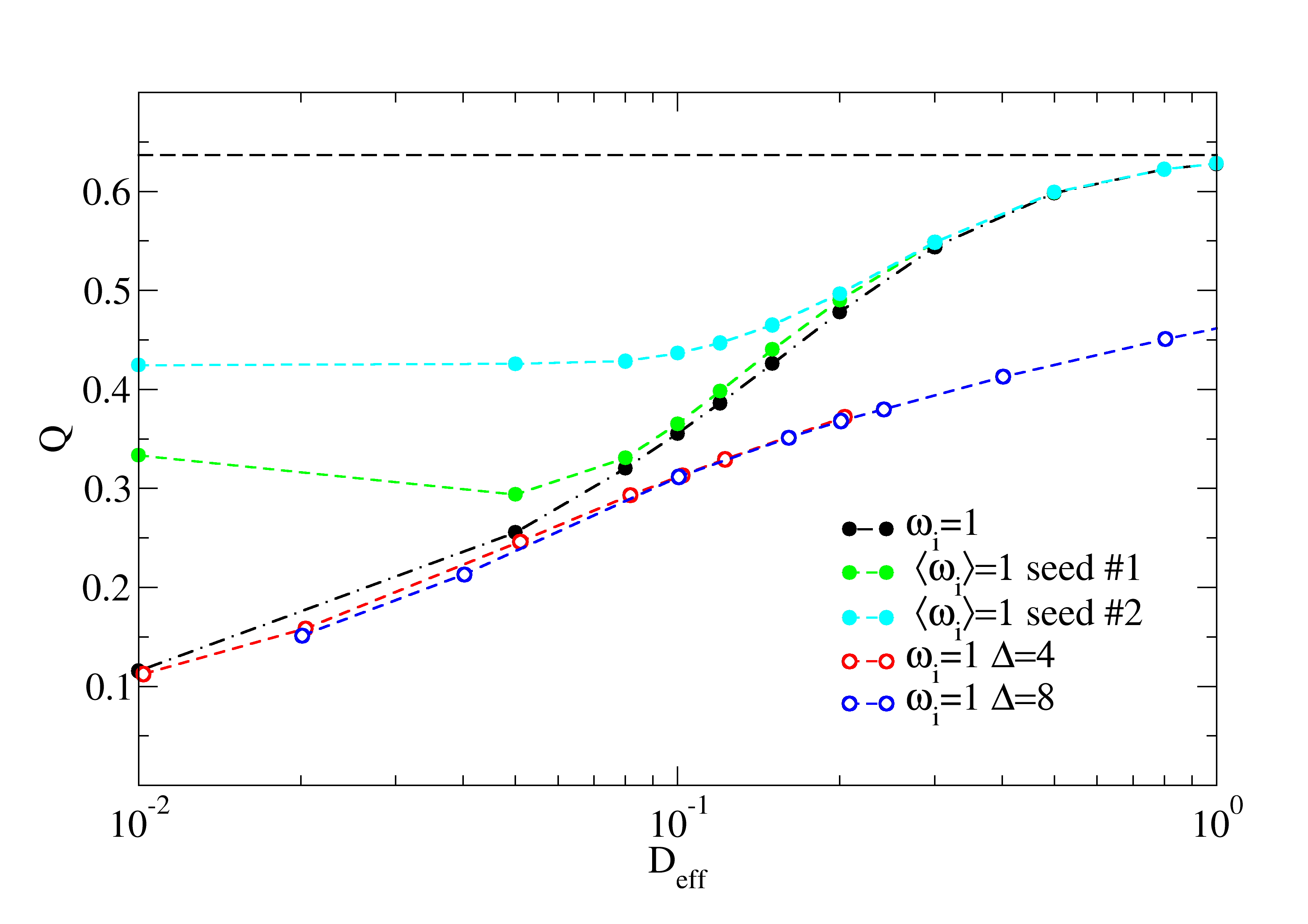}
    \caption{ \label{fig:indnoise_Q} Short range order measured through an order parameters as a function of $D_{eff}$. The plot includes also values for the model with spatially correlated noises discussed in Sec.~\ref{sec:cor}, where $D_{eff}$ depends upon $D$ and the spatial correlation length $\Delta$. In the uncorrelated noise case, instead, $D_{eff}=D$. Small values of $Q$ indicate higher order, while the maximum possible value is shown with a dashed horizontal line. In all cases $k=1$.
    }
\end{figure}

The evidence of short-range order in the chain is shown in Fig.~\ref{fig:indnoise_Q} where the short-range order parameters is analysed, defined as
\begin{equation}
        Q =  \frac{1}{N-1}\sum_{i=1}^{N-1}|\sin(\theta_{i+1}-\theta_i)|
%        Q_2 &= \frac{1}{N-1}\sum_{i=1}^{N-1}\frac{|\mod(\theta_{i+1}-\theta_i,2 \pi)|}{2\pi}
\end{equation}
Small values of $Q$ indicate higher order. $Q$  smoothly increases with $D$, signalling a decreasing order, as expected, with the exception of the case with non-uniform driving, where the order parameter is somehow constant (or slightly decreasing with $D$) at small values of $D$. 
%$Q_2$ displays a sharper transition from order to disorder, in all cases: however it still marks a difference at small $D$ between the uniform and non-uniform driving, the latter being less ordered than the former. 
Somehow the distribution of $\omega_i$ acts as a kind of noise, disturbing the short range order. 
%At large $D$ on the contrary, a distribution in $\omega_i$ slightly reduces $Q_2$, but an explanation for this is beyond our immediate scope. 
It is also clear that a non uniform driving can produce quite different values of $Q$, from sample to sample, at least at small/intermediate values of $D$. 

%%%%%%%%%%%%%%%%%%%%%%%%%%%%%%
\section{Results with correlated noise.} 
\label{sec:cor}

\begin{figure} 
    \includegraphics[width=\columnwidth]{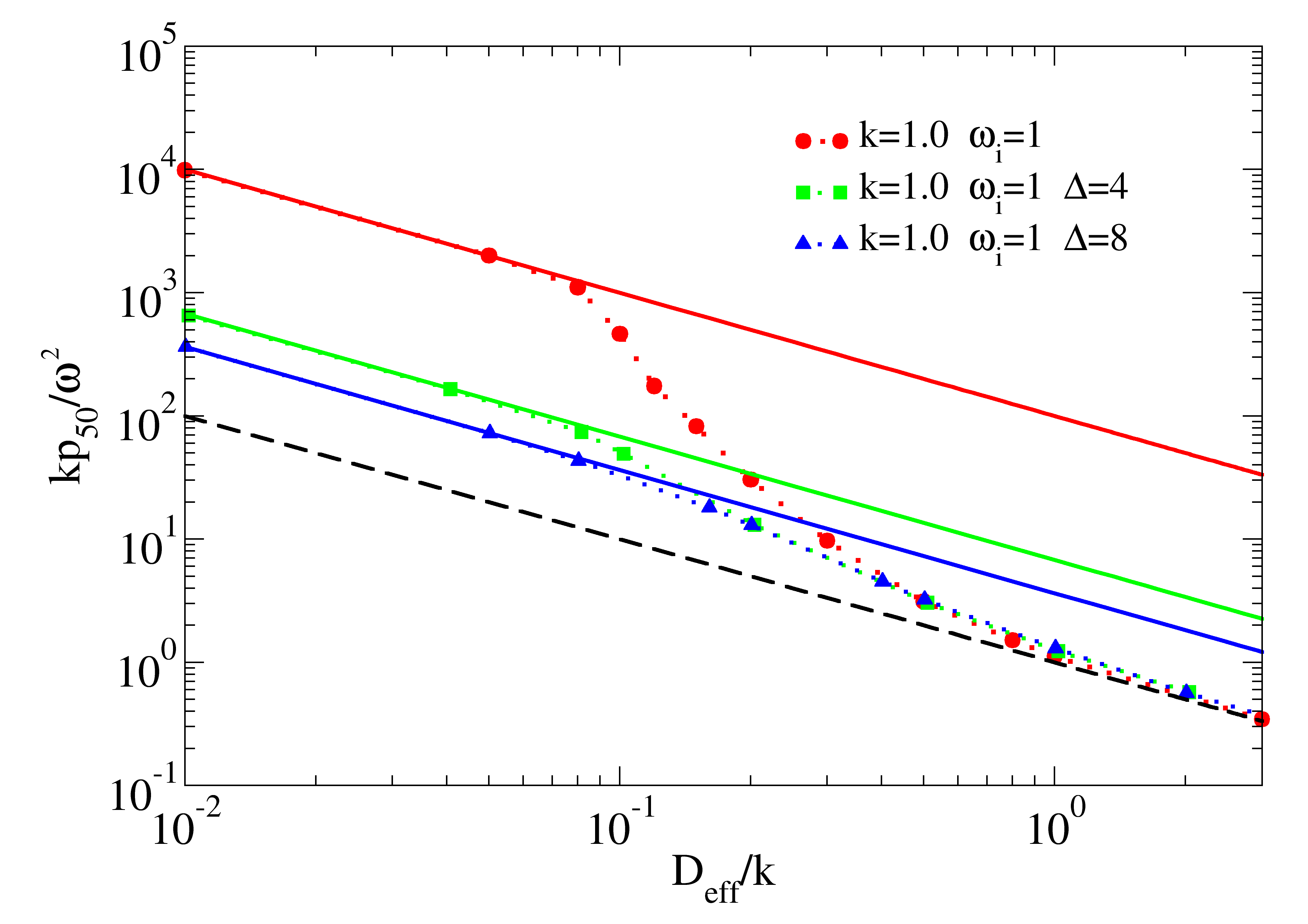}
    \caption{ \label{fig:cornoise} Rescaled precision as a function of $D_{eff}/k$, see text for the definition. The dashed line marks the behavior $k/D_{eff}$, while the solid lines mark the behavior $k/D_{cm}$ with $D_{cm}$ the center of mass diffusivity which is different for each choice of $\Delta$, see discussion in the text.}
\end{figure}

\begin{figure} 
    \includegraphics[width=\columnwidth]{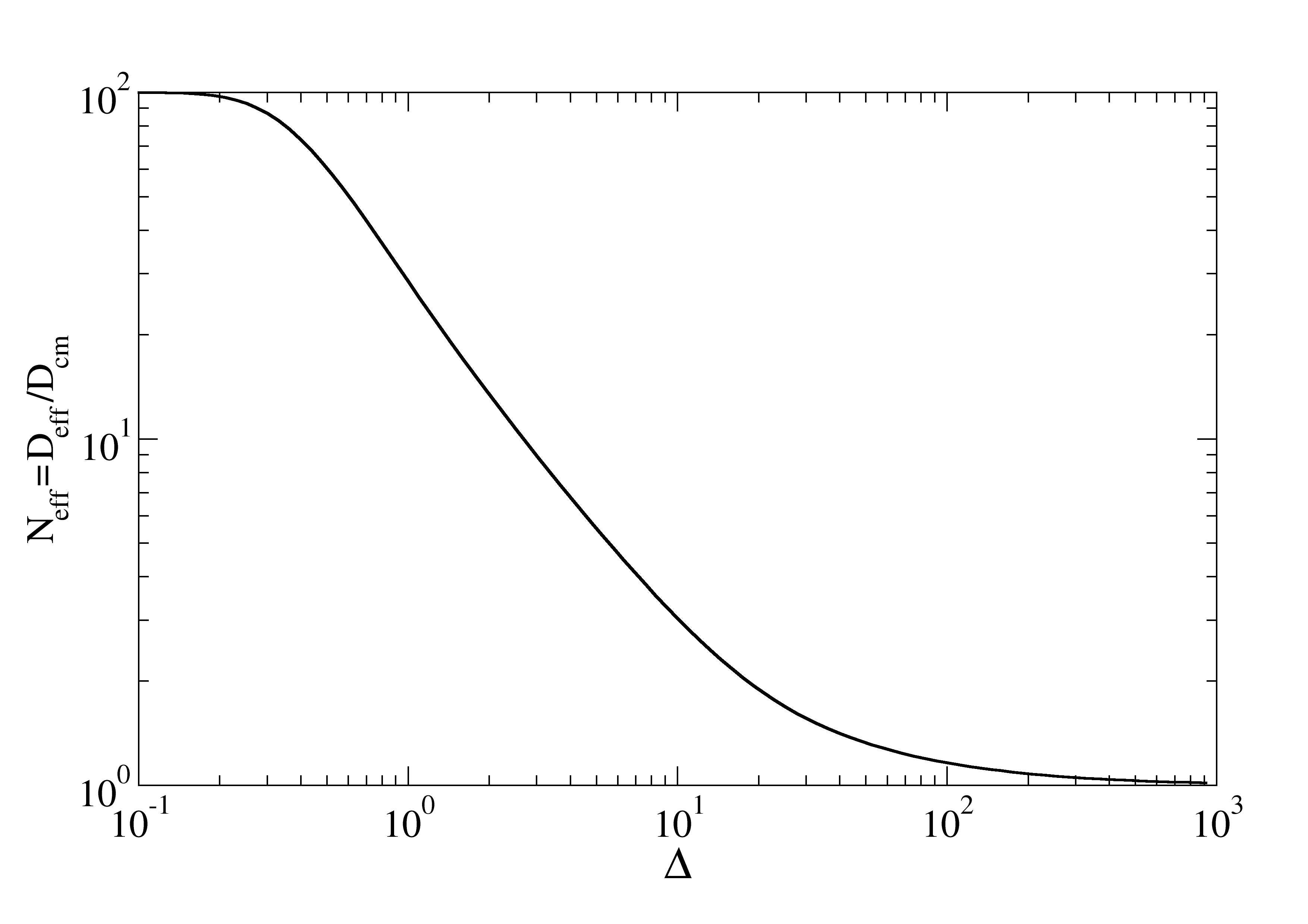}
    \caption{ \label{fig:Dcm} Effective number of independent degrees of freedom expressed  as the ratio $N_{eff}=D_{eff}/D_{cm}$, in presence of spatially correlated noise, as a function of the noise correlation length $\Delta$. When $\Delta \ll 1$ the noise is basically uncorrelated and the effective number of independent degrees of freedom is $N=10^2$. When - on the contrary - $\Delta \ll N$ then the noise correlation spans the entire system and the effective number of degrees of freedom tends to $1$. }
\end{figure}

In order to get noises with space correlation as given in Eq.~\eqref{eq:cor}, we define
\begin{equation}
    \eta_i(t)=\sum_{j=1}^N e^{-\frac{|i-j|}{\Delta}}\xi_j(t)
    \label{correlated_noise}
\end{equation}
where $\xi_j(t)$ are independent white noises with amplitude $1$. As discussed below Eq.~\eqref{eq:cor}, $\Delta$ acts as an effective spatial correlation length for the noise. We note that the auto-correlation $\langle \eta_i(t)^2 \rangle$ of this kind of noise is not $1$, but depends on both $i$ (weakly) and $\Delta$. Therefore, in order to compare the results of such a noise implementation with those in the previous section, we define $D_{eff}=D  \langle \eta_i(t)^2 \rangle$ with $i=50$ the site where we are measuring the precision. With this definition, we plot in Fig.~\ref{fig:cornoise} the precision as a function of $D_{eff}/k$. The idea of this re-scaling of $D$ is that we have on the $x$ axis a measure of the amount (amplitude) of noise on the elected site, and it is interesting to check if this amount is meaningful in some regime. 
In fact, we see that when $D_{eff}/k$ is large enough (on values comparable to the case where the same transition happens for independent noises, Fig.~\ref{fig:indnoise}), any effect of coupling disappears and the precision is that of an uncoupled particle with diffusivity $D_{eff}$, i.e. $p \sim 1/D_{eff}$. On the contrary, when $D_{eff}/k$ is small, the precision saturates on a line which is smaller than $N/D_{eff}$. In Fig.~\ref{fig:cornoise} we plot the lines corresponding to the precision of the center of mass, which is $p_{cm}=\omega^2/D_{cm}$ with $D_{cm}$ discussed in the Appendix. It is seen that one can define an effective number of degrees of freedom $N_{eff}=D_{eff}/D_{cm} < N$ which decreases when $\Delta$ increases (see Fig.~\ref{fig:Dcm}).
Summarising the results in Fig.~\ref{fig:cornoise}, the effect of spatial correlations in the noise is negligible for the precision at strong noise, but becomes relevant and reduces $p$ at small noise. Interestingly, the effect of increasing $\Delta$ on the short-range order of the phase, when $D_{eff}/k$ is small, is negligible, see Fig.~\ref{fig:indnoise_Q}. For larger values of $D_{eff}/k$ some ordering effect is seen but it is less clear because it depends upon the choice of the observable. 
\begin{figure} 
    \includegraphics[width=\columnwidth]{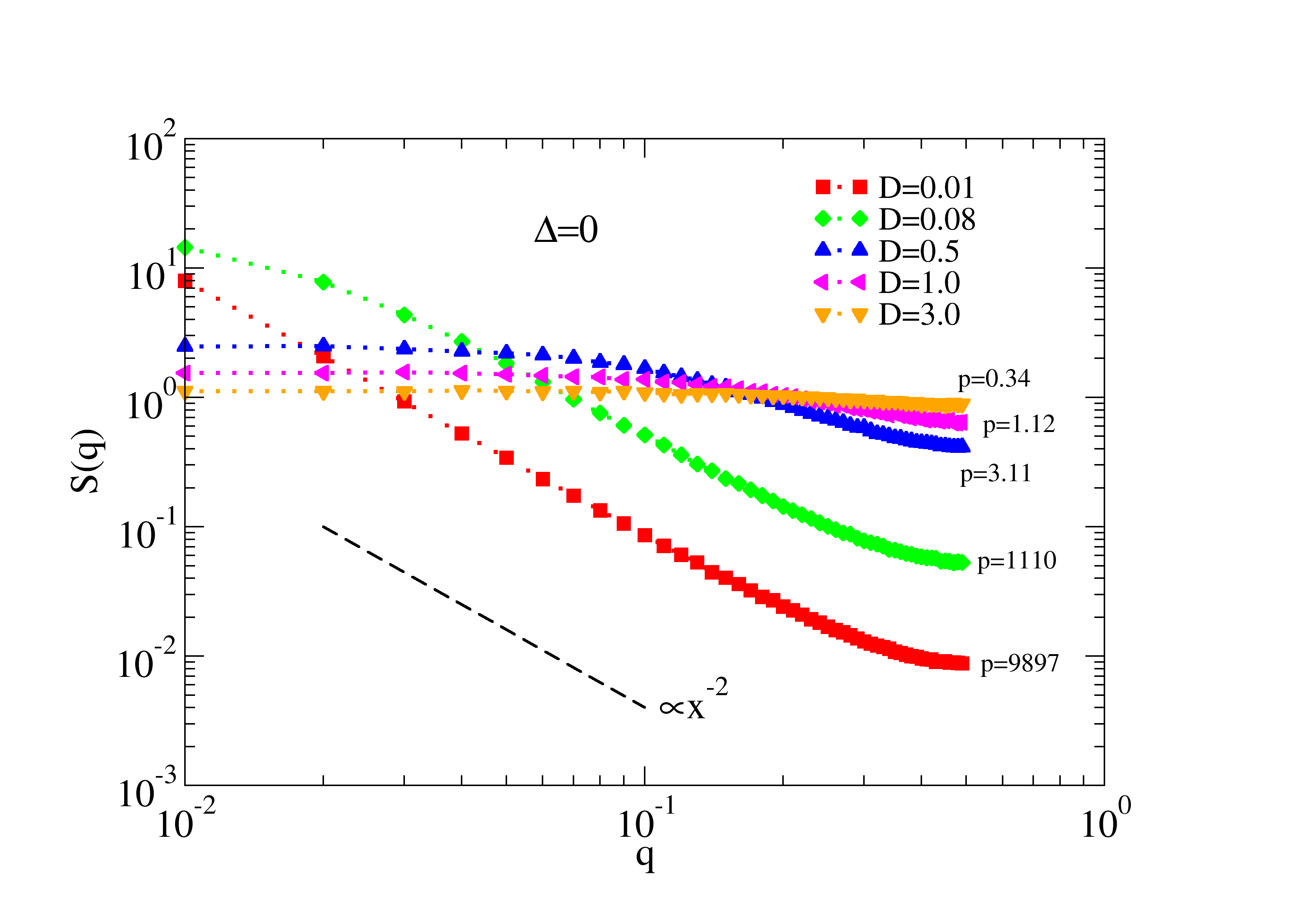}
    \includegraphics[width=\columnwidth]{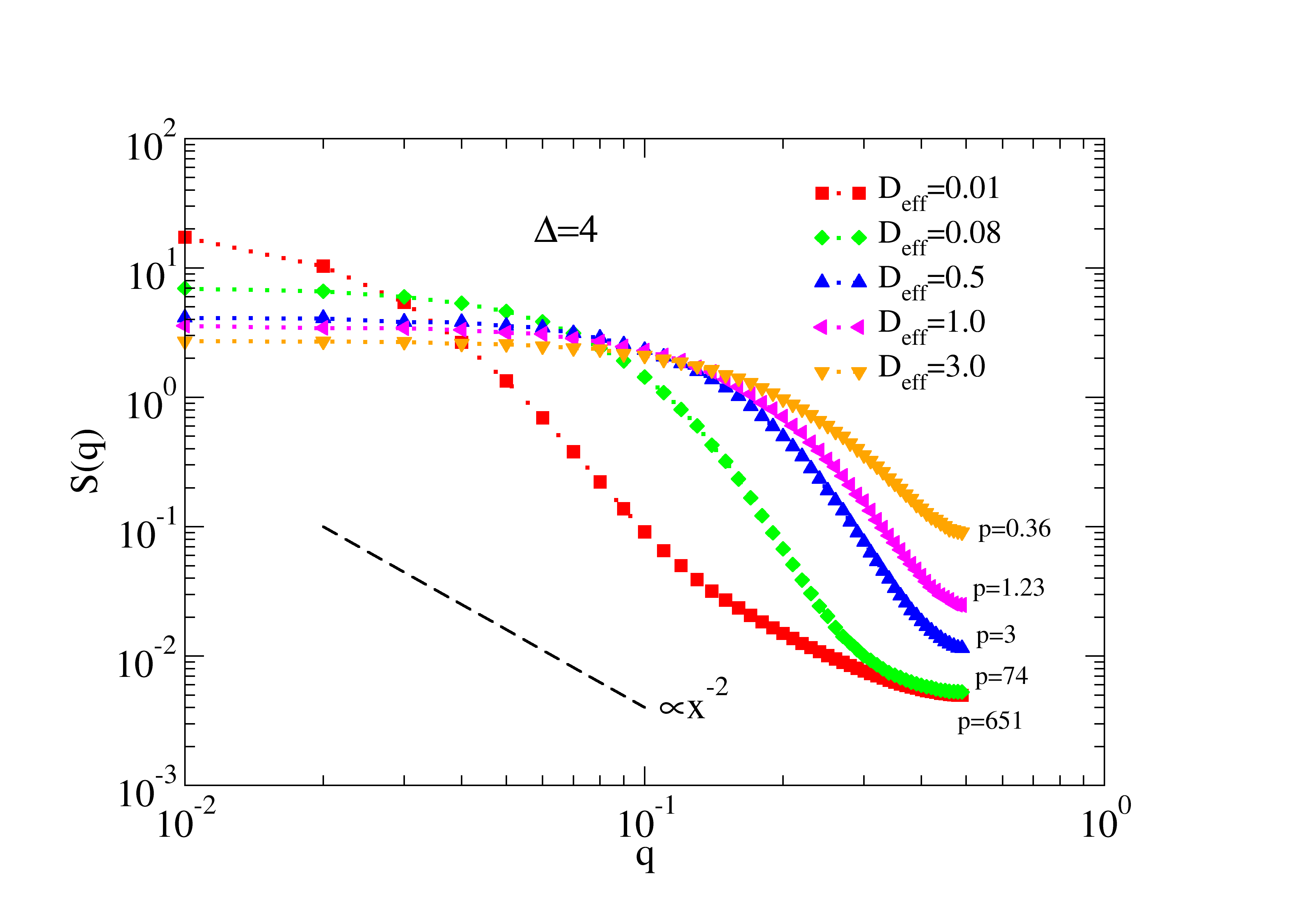}
    \includegraphics[width=\columnwidth]{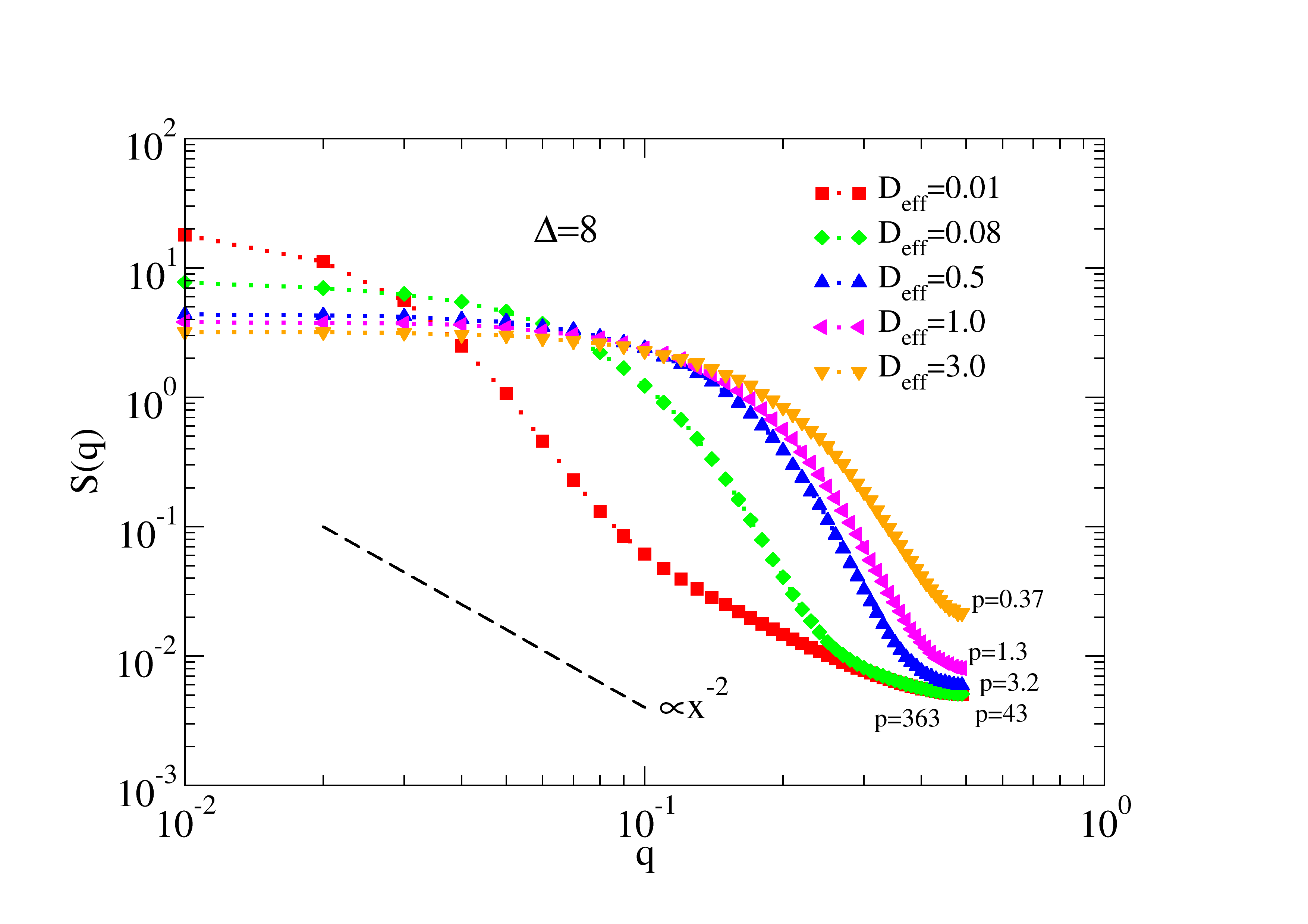}
    \caption{Power spectra of the field $\theta_i$ for several choices of $\Delta$ and $D$ or $D_{eff}$. For each curve the corresponding value of the precision $p$ for the site $i=50$ is shown. \label{fig:spectra}}
\end{figure}
To get some further insight in the question of spatial order in the chain and of its dependence with $\Delta$, we plot in Fig.~\ref{fig:spectra} the power spectra of the field $\sin \theta_i(t)$. Those are defined as 
\begin{equation}
    S(q)=\frac{2}{N} \left\langle \left| \sum_{i=1}^N e^{-2\pi\mathcal{I} q i/N} \sin{\theta_i(t)} \right|^2 \right\rangle
\end{equation}
where $\mathcal{I}$ is the imaginary unit.
The case $\Delta=0$ is the most clear one, it goes from an almost flat spectrum at large $D$, with very small precision, to a perfectly $q^{-2}$ behavior marking high spatial order and very large precision. When $\Delta>0$ we observe two main differences: 1) the spectrum is non-flat (i.e. there is spatial order in the phase) even at very large $D_{eff}$, this is remarkable if compared with the fact that the precision is the same as in the case of $\Delta=0$ i.e. it is the uncoupled value $\sim 1/D_{eff}$; 2) an additional lengthscale, certainly related to both $D_{eff}$ and $\Delta$, appears at small values of $D_{eff}$, representing a maximum length below which the $q^{-2}$ behavior is observed and above which is lost.

%%%%%%%%%%%%%%%%%%%%%%%%%%%%%%
\section{ Conclusions} 
\label{sec:concl}

In conclusions, we have put in evidence the non-trivial effect - on the precision - of different sources of correlation in a chain of coupled phases. When the coupling  is induced by a simple  force aligning the phases of  adjacent sites, the precision fairly reflects the degree of order in the phase field, i.e. it is larger when the spatial order is higher. On the contrary, when there is a correlation in the noises, even if the degree of order in the phase field is not strongly influenced, the precision dramatically drops as if the number of effective degrees of freedom is reduced. We can summarise our findings in the following way: the precision $p$, being influenced by the diffusion coefficient, reflects a degree of dynamical order which is not entirely explained by the degree of spatial (static) order.

Our observations represent, in our opinion, a first step in order to understand the low precision observed in the fluctuations of the flagellar beating such as for the tail of sperms and for the cilia of C. reinhardtii. Certainly the connection with biophysics must go through more refined and realistic models of noisy flagellar beating, whose investigation is in progress.

\begin{acknowledgments}
The authors would like to thank Claudio Maggi and Alessandro Sarracino for useful discussions.
\end{acknowledgments}

\appendix
\section{Spatial correlation of noise}
In order to obtain the expression in Eq.~\eqref{eq:cor} we need to evaluate the various contributes for the different values of the quantities in the exponents. Considering, first, the case with $j>i$, we can split the sum into three parts depending on the $j$ values, that is
\begin{equation}
\begin{split}
  \langle \eta_i(t)\eta_j(t)\rangle =&\sum_{n=1}^{i-1} e^{-(i+j-2n)/\Delta}+ \sum_{n=i}^{j-1} e^{-(j-i)/\Delta} +\\
  &\sum_{n=j}^{N} e^{-(2n-i-j)/\Delta}.
\end{split}  
  \label{eq:sumsplit}
\end{equation}
Using the expression for the sum of a geometric series, the terms in Eq.~\eqref{eq:sumsplit} can be written as
\begin{equation}
  \begin{split}
  \langle \eta_i\eta_j\rangle =& \frac{e^{-(j+i)/\Delta}}{1-e^{-2/\Delta}} \Big( e^{2(i-1)/\Delta} -1 \Big) +\\
  & (j-i)e^{-(j-i)/\Delta}+ 
  \frac{e^{(j+i)/\Delta}}{1-e^{-2/\Delta}} \Big( e^{-2j/\Delta}- \\ & e^{-2(N+1)/\Delta}\Big) = 
  \frac{1}{1-e^{-2/\Delta}}\Big \{ e^{-(j-i)/\Delta} \cdot \\
  &\Big [j-i+1+
   (1-j+i)e^{-2/\Delta}\Big ] \\
   &-e^{-(j+i)/\Delta} - e^{-(2N+2-j-i)/\Delta}\Big \}.
\end{split}  
  \label{eq:noisecorr}
\end{equation}

The final expression in Eq.~\eqref{eq:cor} is obtained taking into account that the case $j<i$ i given by exchanging the indices and using the definitions $d\equiv j-i$ and $s\equiv i+j$.\\

Putting $j=i$ in Eq.~\eqref{eq:noisecorr} we obtain the self-correlation of the noise $\eta_i$ that is given by

\begin{equation}
\langle \eta^2_i\rangle=    \frac{1+e^{-2/\Delta}-e^{-2i/\Delta} -
  e^{-2(N+1-i)/\Delta}}{1-e^{-2/\Delta}},
\end{equation}

that it can be used to have the quantity $D_{eff}=D\langle \eta^2_i\rangle$.\\
Considering the Eqs.~\eqref{eq:cm} and \eqref{correlated_noise} the noise of the average motion can be written as

\begin{equation}
\eta_{cm}(t)=\sum_{i=1}^{N}\eta_i(t)=
\sum_{i=1}^{N}\sum_{j=1}^{N} e^{-|i-j|/\Delta} \xi(t)
\end{equation}

and considering the properties of noise $\xi(t)$ its amplitude is 

\begin{equation}
\langle \eta_{cm}^2\rangle=\sum_{i=1}^{N}\sum_{j=1}^{N} \sum_{n=1}^{N} 
e^{-|i-j|/\Delta} e^{-|n-j|/\Delta}.
\label{eq:cmnoise}
\end{equation}
The final expression of $\langle \eta^2(t)\rangle$ can be obtained evaluating the different contributes of the absolute values and, after long but simple calculations, we can write that

\begin{equation}
\begin{split}
    \langle \eta_{cm}^2\rangle=&\frac{1}{(1+e^{-1/\Delta})(1-e^{-1/\Delta})^3} \Big\{ N(1-e^{-2/\Delta})\cdot \\
    &\Big [1+e^{-2/\Delta}+2e^{-1/\Delta}\Big ( 1+e^{-N/\Delta} \Big )\Big ] -4e^{-1/\Delta}-\\
    &6e^{-2/\Delta}-4e^{-3/\Delta}+4e^{-(N+1)/\Delta} +
    8e^{-(N+2)/\Delta}+ \\
    &4e^{-(N+3)/\Delta}-2e^{-2(N+1)/\Delta} \Big \}.
\end{split}
\label{eq:cmnoisefin}
\end{equation}
The $D_{cm}$ value in the main text is given by $D_{cm}=D\langle \eta_{cm}^2\rangle/N^2$.
\bibliographystyle{apsrev4-1}
%\bibliography{biblio,example,biblioprr}
\bibliography{biblio,biblioprr,spermprecision2}

%\bibliography{biblio}% Produces the bibliography via BibTeX.
\end{document}